\documentstyle[prl,aps,epsfig,preprint,amsfonts,amssymb,amsmath,rotate] {revtex}
\tightenlines
\begin{document}
\advance\textheight by 0.2in
\draft
\title{The Architecture of Idiotypic Networks: Percolation and Scaling Behaviour}
\author{Markus Brede and Ulrich Behn}
\address{Institut f\"ur Theoretische Physik, Universit\"at Leipzig, 
Augustusplatz 10, 
 D-04109 Leipzig, Germany}
\maketitle
\begin{abstract}
We investigate a model where idiotypes (characterizing B-lymphocytes and antibodies of an immune
system) and anti-idiotypes are represented by
complementary bitstrings of a given length $d$ allowing for a number of
mismatches (matching rules). In this model, the vertices of the hypercube in
dimension $d$ represent the potential repertoire of idiotypes. A random set
of
(with probability $p$) occupied vertices corresponds to the expressed
repertoire of idiotypes at a given moment.  Vertices of this set linked by the above matching rules build random clusters.
We give a structural and statistical characterization of these clusters --
or in other words -- of the architecture of the idiotypic network. Increasing
the probability $p$ one finds at
a
critical $p$ a percolation transition where for the first time a large
connected graph occures with probability one. Increasing $p$ further, there
is
a second transition above which the repertoire is complete in the sense
that any newly introduced idiotype finds a complementary anti-idiotype. We
introduce structural characteristics such as the mass distributions and
the fragmentation rate for random clusters, and determine the scaling
behaviour of the cluster size distribution near the percolation transition,
including finite size corrections. We find that slightly above the
percolation transition the large connected cluster (the central part of the
idiotypic network) consists typically of one highly connected part and a
number of weakly connected constituents and coexists with a number of small,
isolated clusters. This is in accordance with the picture of a central and a
peripheral part of the idiotypic network and gives some support to idealized
architectures of the central part used in recent dynamical mean field
models.

\end{abstract}

\pacs{{PACS numbers: 64.60.Ak, 05.10.Ln, 02.70.Lq, 87.18.-h}}


\vspace*{0.5cm}

\section{Introduction}

B-lymphocytes carry on their surface highly specific receptors, so-called antibodies. If these
receptors detect complementary structures, the lymphocyte is stimulated to proliferate and after
several generations and differentation into plasma cells secretes antibodies of the same specificity.
B-lymphocytes and antibodies of a given specificity are said to have a certain idiotype.
Complementary structures to an idiotype are antigen or other, anti-idiotypic antibodies. Between
B-lymphocytes of different idiotype thus emerges a functional network of mutual stimulation and
inhibition, the idiotypic network \cite{NKJ}. The idiotypic network is supposed to at least
partially contribute to the functionality of the immune system as, e.g., immunological memory or suppression of
autoreactive clones. Though quantitative data are very hard to access by experiment there are
some recent observations which underline the importance of idiotypic interactions \cite{p53,IICrev}.

New idiotypes are produced in the bone marrow or due to hypermutation during the proliferation of
stimulated lymphocytes which introduces a random metadynamics of the repertoire. At a given moment
the random network has a certain architecture. The aim of this paper is to give a statistical
description of this architecture. Knowledge of the typical architecture of the
idiotypic network is crucial for describing the population dynamics of the interacting B-lymphocytes
and antibodies, cf. \cite{RAP,TOP}, which is however not the subject of this paper.

Derived from hypotheses of theoretical immunology (cf. \cite{NKJ,Co1}, for a recent review see \cite{Ueb}) we present a statistical
analysis of bit-string based networks, which shows, that this approach is well suited to reproduce
reasonable network structures. Especially, our considerations show, that realistic network
topologies can be conceived as an extension of first approaches to that problem which assumed a
Bethe-lattice structure. Suggestions of how loops should be added to such structures have been made
previously \cite{NW92b}. For this, our work provides a very natural access.

Generally, idiotypic networks are supposed to realize a tradeoff between two basic requirements: they should
contain a great number of small isolated components, but on the other hand still be able to respond to
arbitrary antigen, that means being complete. Small components are thought necessary to store information about previously encountered antigen
\cite{Ueb,KL1,KL2}.
The existence of such components does obviously demand a low connectivity of the network. With
completeness on the other hand it is assumed that of a great number of antibodies each is able to detect
many different types of antigen. Hence the network connectivity should not be too low.

With regard to the underlying biological problem the one great--many small clusters situation is thus
worthwhile to earn special attention. Theoretical immunologists suppose that the idiotypic
network should consist of a large number of small clusters. On the other hand, as a consequence of
relatively high connectivities, also a large component should be contained within the idiotypic network,
which is denoted as its central part. This great cluster could play an important role in the control of
autoreactive clones \cite{Co1,CogPriII,SecGen,B}.

In the following the bit-string model of \cite{Farmer} will be explained briefly. Basically, antibodies are
identified with bit chains of a given length $d$. Thus, there are $2^d$ different antibody types. The set
of all conceivable antibodies, i.e. the potential repertoire, is then represented by $\{ i=(i_1,...,i_d),
i_j\in \{0,1 \} \}$. Estimating the probable size of networks that can be complete in the above sense shows
that $d \approx 32...36$ should be a good value for realistic models \cite{PeOs}.

Antibodies recognize each other if they have complementary structures, i.e. if they are represented by
perfectly complementary bit strings. If there are small deviations from the exact structural
complementarity a matching is still possible, though with lower affinity. This is described by
so-called matching rules.

For example we imagine that antibodies react if the respective structures are complementary except one
small region. Using the language of our model the corresponding bit-strings should be inverse except one
bit that belongs to the non-complementary region. We call this kind of rule a one-mismatch rule.
Two-mismatch rules, which allow reactions between antibodies that are complementary up to two
non-matching regions, are accordingly defined by connecting bit-strings that are complementary with two
exceptions allowed. Naturally, the multitude of possible rules is not exhausted by mismatch rules.
Rules that express matching of mutually shifted antibody parts could be conceived as well. For the sake of
simplicity this work is confined to those matching rules that are associated with reactions of highest
affinity, i.e. inversion, one- and two-mismatch rules.

\begin{figure}[tbp] 
 \begin{center}
  \epsffile{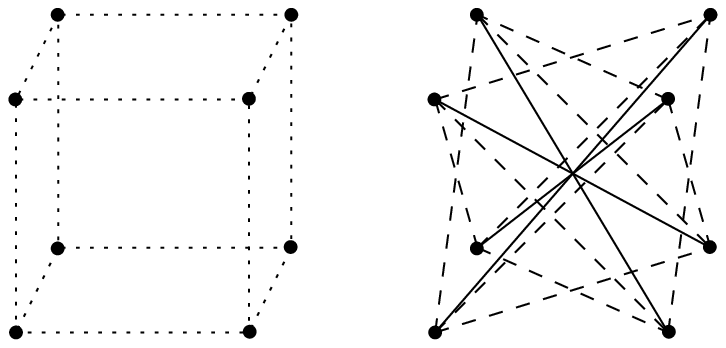}  
  \vspace*{0.5cm}
  \caption[soso]{Comparison of a hypercube (left side) with the one-mismatch base graph (right
  side) which have the same set of vertices ($d=3$). Solid lines connect perfectly complementary
  vertices, dashed lines mean one-mismatch links.}
  \label{cube}
 \end{center}
\end{figure}

\begin{figure}[tbp]
  \begin{center}  
  \epsffile{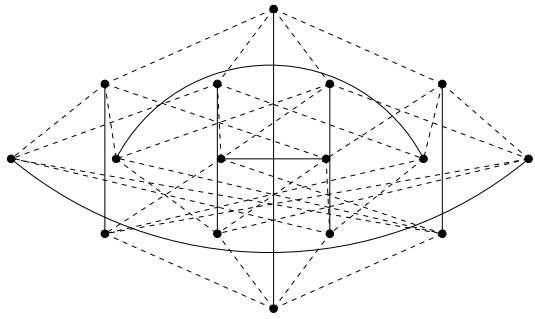} 
  \vspace*{0.5cm}
  \caption[soso]{One-mismatch base graph in $d=4$. Same notation as in Fig. \ref{cube}.}
  \label{smallgraph}
 \end{center}
\end{figure}

Mathematically, the set of all antibody types (represented by bit-strings or vertices of a hypercube) together with
all possible reactions between them defines a graph. Thus, we want to call
\begin{align} \nonumber G^1_d=&( \{
v=(v_1,...,v_d), v_j \in \{0,1\} \}, \{ v \textrm{ connected with $w$}\\ \nonumber &\textrm{ if } v_i=\overline{w_j}
\textrm { for all } i,j=1,...,d \textrm{ except one position} \\  &\textrm{ at the most} \} ) 
\end{align} the
one-mismatch base graph and
\begin{align} \nonumber G^2_d=&( \{ v=(v_1,...,v_d), v_j \in \{0,1\} \}, \{v \textrm{
connected with $w$}\\ \nonumber & \textrm{if } v_i=\overline{w_j} \textrm { for all } i,j=1,...,d \textrm{ except two
positions} \\  &\textrm{ at the most} \})
\end{align} the two-mismatch base graph, respectively. To obtain a better
impression of one-mismatch graphs they can be compared to hypercubes of the same dimension. Looking, e.g., on
hypercubes of dimension $d=3$ edges of one-mismatch graphs are represented by all space and side
diagonals. This, together with a picture of an one-mismatch graph in dimension $4$ is shown in Figs. \ref{cube} and \ref{smallgraph},
respectively.

The actual repertoire, i.e. the set of all types of idiotypes that are
really existing within the body at a given time, is a subset of the
potential repertoire.
There are profound reasons to believe that the actual repertoire should be distributed randomly within the potential repertoire
\cite{Ueb}. A way of realizing this is to choose antibodies with a certain probability $p$. Then the idiotypic network
is represented by the random graph that is composed of all occupied vertices (antibodies) and their connections
(possible reactions). Clearly, this is a site percolation problem \cite{AhSt}.

Percolation problems on standard lattices are far away from beeing a new field of research. Percolation on one-mismatch graphs,
however, bears some major difference to ordinary percolation.

Percolation is always connected to an abrupt change of some system property (the standard property
is the existence of a connecting path from the upper to the lower boundary on 2-dimensional lattices) if the occupation
probability passes a certain value, the percolation threshold. This transition becomes sharp in the limit of
infinite systems.

On lattices there is no question how to increase the system size to infinity. However, sequences of
one-mismatch graphs exhibit relations between small and large systems that are essentially different from similar
characteristics of lattices which are created by the multiplication of a unit cell. Understanding of
these distinctions comes from studying the structure of fully occupied one-mismatch graphs (base graphs) which will be
performed in the ensuing Section.

In this paper we investigate the architecture of functional networks built
by constituents of randomly generated characteristics interacting with
complementary constituents. Our motivation, the formulation of the problem
and the interpretation of the results are all in the context of modeling the
immune system using language and methods of statistical physics. However,
our results are essentially independent of the specific immunological
interpretation and could be of broader interest. The situation of
interacting randomly generated complementary constituents is quite general.
For example think of chemical reactions (origin of life), ecosystems, social
networks.

Generally, the conception of the paper is as follows: in Sec. II topological properties of base-graphs will be
studied. Section III gives an analysis of the underlying percolation problem, which then leads, together with
applications of random graph theory from Sec. IV and a study of intrinsic structures of great clusters in Sec. V, to
conclusions about parameter regimes, in which bit-string model induced random graphs resemble idiotypic
networks. In the appendix another method to calculate thresholds via the renormalization of small
cells will be discussed.

\section{Structure of the Base Graph for the One-Mismatch Rule} The above definition of matching rules allows an
easy calculation of distances between vertices of $G^1_d$. Let $i,j \in G^1_d$ be vertices, $d_H$ their Hamming
distance (i.e. the number of different bits between them) and
\begin{equation}
 \label{metric}
 d_G(i,j)=\inf_{ \begin{substack} {\textrm{pathes $w$ in $G^1_d$} \\ \textrm{ connecting $i$ and $j$}} \end{substack}} l(w) \textrm{
,}
\end{equation} where $l(w)$ denotes the length of a path $w$, a metrics. Then it holds that if
\begin{align}
 \nonumber
  &d_H<d/2 &d_G &= \begin{cases} d_H   & \hspace*{0.65cm} \textrm{for $d_H$ even } \\
                                 d_H+1 & \hspace*{0.65cm}  \textrm{for $d_H$ odd } \\
		   \end{cases} \\ \label{eq1}  
  &d_H=d/2 & d_G&=d/2 \hspace*{1.7cm} \textrm{ for $d$ even }\\ \nonumber 
  &d_H>d/2 &d_G &= \begin{cases} d-d_H+1 &\textrm{for $d-d_H$ even } \\
                                 d-d_H   &\textrm{for $d-d_H$ odd , } \\
		   \end{cases}  
\end{align} which shows that the maximum distance $D_{\textrm{max}}(d)$ between
vertices is roughly the half of that on ordinary hypercubes. This is essentially due to the inversion rule.
Furthermore, it should be noticed that $D_{\textrm{max}}(d)$ does not change when increasing the dimension by one from odd
($d=2n+1$) to even ($d=2n+2$) values.

To obtain additional information about the toplogy we distinguish vertices corresponding to their distance $s$ to an arbitrarily chosen origin, e.g. $v=0=(0,...,0)$, whose choice is arbitrary due to the specific construction-rule of base graphs. We denote the set of vertices with distance $d_G(v,i)$ from $v$ by
\begin{align}
 E_s = \{ i\in G^1_d| d_{G}(v,i)=s \} \label{eq2} \textrm{.}
\end{align}
Applying (\ref{eq1}) it is possible to compute the number of vertices $|E_s|$ belonging to $E_s$ (mass distribution) by simple combinatorics to
\begin{align}
 |E_s|=
 \begin{cases}
   \begin{pmatrix}
    d \\ d_{G}-1
   \end{pmatrix}
   & \textrm{ if } d_{G}= \frac{d+1}{2}
   \\
   \\
   \begin{pmatrix}
    d+1 \\ d_{G}
   \end{pmatrix}
   & \textrm{ if } d_{G}\leq \frac{d}{2}
 \end{cases}
 \textrm{.}
 \label{eq3}
\end{align}

\begin{figure}[tbp]
  \begin{center} 
  \epsfysize 5 cm
  \epsfxsize 8 cm
  \epsffile{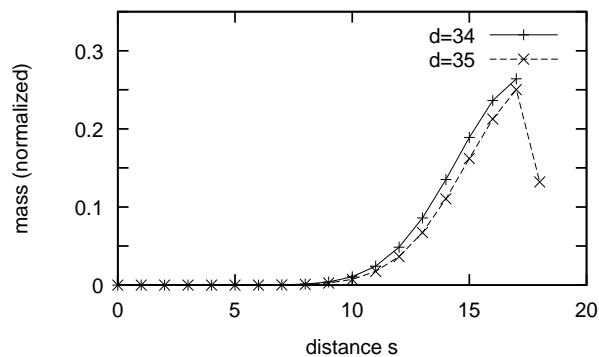}     \vspace*{0.5cm}
  \caption[Massenverteilung]{Normalized mass in the dimensionless distance $s$ from an arbitrary origin in even ($d=34$) and odd ($d=35$) dimension. Points have been connected to guide the eyes.}
  \label{pic1}
 \end{center}
\end{figure}

A visualization of Eq. (\ref{eq3}) in Fig. \ref{pic1} shows a major difference between odd and even dimensions $d$.
Generally, $|E_s|$ grows monotonically with increasing distance. For odd dimensions, however, the number of vertices
with maximal distance $D_{\textrm{max}}$ from $v$ is substantially smaller than that with distance
$D_{\textrm{max}}-1$

Differences in the overall structure of graphs in odd and even dimensions $d$ become even more obvious if the number of links connecting vertices of the same set $E_s$ is considered. Using (\ref{eq1}), a somewhat lengthy but straightforward calculation shows that connected vertices in the same distance $s$ from the origin $v$ occur in
\begin{align}
  \nonumber
  &\textrm{$d$ even only for } &&s=D_{\textrm{max}}=d/2 \\ \label{eq4}
  &\textrm{$d$ odd only for } &&s=D_{\textrm{max}}=(d+1)/2 \textrm { or } \\ \nonumber
              &                &&s=D_{\textrm{max}}-1 \textrm {.} 
\end{align}
Analogously, the number of links that connect vertices of $E_s$ with vertices belonging to $E_{s+1}$
may be computed to $d+1-s$ for $s<D_{\textrm{max}}-1$.

As a consequence there must exist loops of even as well as of odd length. Moreover, all short loops with
length less than $d$ must be even (since only loops comprising links between vertices in the same distance from any
loop element can be of an odd length).

An important property of standard systems in percolation theory is that the smaller system is always contained
within the larger one. For one-mismatch graphs, however, this is not possible. Let $d_1<d_2$ be dimensions of
$G^1_{d_1}$ and $G^1_{d_2}$, respectively. If $G^1_{d_1}$ were contained in $G^1_{d_2}$ it would follow directly
that $G^1_{d_2}$ should have uneven loops with length less then $d_1$. This gives a contradiction since $G^1_{d_2}$
contains no odd loops smaller than $d_2$, whereas $d_1<d_2$ by assumption. 

Resuming these results we state that percolation on one mismatch graphs differs from standard percolation. A crucial
point for this distinction is the way how global system properties change if the system size approaches infinity.
Nevertheless, methods of percolation theory can be applied to this
kind of problem. Even more, choosing an appropriate majority rule renormalization group procedures can be extended to
one-mismatch-graphs (see appendix A).

To the best of our present knowledge percolation problems on graphs with similar properties have not been dealt with. Even
percolation on regular hypercubes has found relatively little attention in the physical literature \cite{Camp}.

\section{Cluster size distribution}

Having collected some simple properties of the underlying base graphs we consider now vertices of that graphs occupied with a given probability $p$. The set of all occupied vertices $\Gamma$ together with all bonds, which connect two vertices belonging to $\Gamma$ forms a random graph. In terms of percolation theory a maximal set of  connected vertices is called cluster.

\begin{figure}[tbp]
 \begin{center}
  \epsffile{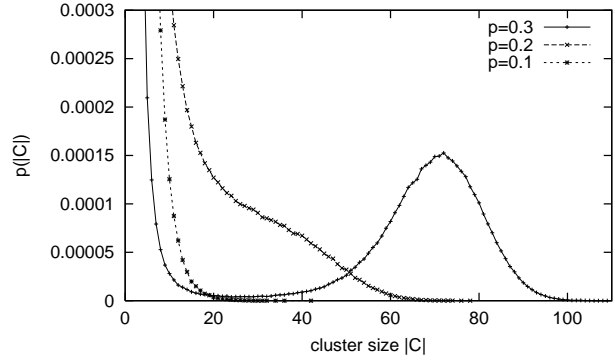}
  \vspace*{0.5cm}
  \caption{Cluster size distribution depending on the occupation probability $p=0.1,0.2,0.3$ for base graphs in $8$ dimensions. The histogram was taken over $10^5$ configurations.}
 \label{picsim1}
 \end{center}
\end{figure}

What is the probability that an arbitrarily chosen vertex belongs to a cluster of size $|C|$? This
question has already been addressed in \cite{KL1,KL2}
where three major regimes of the system have been identified (see also the numerical results in Fig.
\ref{picsim1}):

(i) The first typical situation arises for small values of the occupation probability $p$. Then only
small clusters (whose individual size doesn't make up a finite fraction of the whole system) are expected to appear.

(ii) Increasing $p$ and approaching infinite dimensions, a sharp transition (in the sequel denoted as percolation
transition) to an one great cluster--many small clusters regime was found. At the percolation transition some characteristics
obey scaling laws. These and an approximation to compute percolation thresholds will be dealt with in Sec. IV.

The one great--many small clusters situation is worthwhile to earn
special attention (cf. Sec. I). The great cluster could play an important role in the control of
autoreactive clones. To fulfill that purpose it is believed to have a certain internal structure
(cf. \cite{SecGen,B,Var}, for experimental data cf. \cite{Ker}),
which will be discussed in more detail in Sec. VI. Thus, as stated
already in \cite{KL1,KL2} the simple one--parametric bit chain model exhibits an interesting similarity to the
idiotypic network.

(iii) Finally, if the occupation probability is further increased, a state where random graphs consist of one
connected component only will be reached. Relying on some general results of the theory of random graphs it can be
shown that this indeed marks a second (in the limit of infinite systems) sharp transition which we call
denseness transition. In Sec. V these considerations are resumed and discussed with reference to the completeness of the
immune system.

Exactly the same two thresholds have been considered earlier for a different class of random graphs \cite{ErdRey,Palmer} and more recently in the context of networks of RNA secondary structures \cite{Rei2,Rei1}.

\section{Scaling laws and Bethe approximation}

A common method to analyze percolation problems is the introduction of perimeter polynomials $D_{|C|}$. Using the standard notation we define
\begin{align}
 p(|C|)=p^{|C|}\sum_{S_{\textrm{free}}(C)}(1-p)^{|S_{\textrm{free}}(C)|}=p^{|C|}D_{|C|}(1-p) \textrm{,}
\end{align}
where $p(|C|)$ denotes the probability that a vertex on the base graph belongs to a cluster of size $|C|$.
$S_{\textrm{free}}(C)$ means the free surface, i.e. the set of all unoccupied vertices of the base graph that are adjacent to vertices of $C$.

Generally, there are few problems which allow the explicit calculation of $D_{|C|}$ for arbitrary cluster size $|C|$. Notwithstanding, it is always possible to compute $D_{|C|}$ for small values of $|C|$. This provides a basis for the application of series expansion techniques.
We investigate the structure of small clusters up to size $|C|=5$ on one-mismatch-graphs. As a result for large base graphs ($d>5$) we find the following perimeter polynomials for one-mismatch-graphs
\begin{alignat}{2}
\label{U1}
D_1(q) &= q^{d+1} \textrm{ ,} \\
\label{U2}
D_2(q) &= (d+1)q^{2d} \textrm{ ,}\\
\label{U3}
D_3(q) &= \frac{3}{2}(d+1)dq^{3d-2} \textrm{ ,} \\
\label{U4} \nonumber
D_4(q) &= q^{4d-4}(2(d-1)d(d+1)+\frac{d(d+1)}{2}+\\ + &\frac{2}{3q}(d-1)d(d+1)) \textrm{ ,}\\
\label{U5} \nonumber
D_5(q) &= \frac{5}{2}q^{5d-6} (d-1)d(d+1)(d-2+\frac{1}{q}(d-1)+ \\ & \frac{1}{q^2}+\frac{1}{12q^3}(d-2) ) \textrm{ ,}
\end{alignat}
where $q=1-p$.

In the following we show that in the limit of high dimensions random graphs on one-mismatch-graphs for small values of $p$ are very similar to random graphs on Bethe-lattices with the same coordination number. Arguments of this kind have already been applied to percolation on hypercubic lattices \cite{LD2} making use of the fact that loops in random graphs on sparsely occupied high dimensional lattices are infrequent. Indeed, the first terms of a high dimension expansion on a hypercubic lattice of dimension $d$ are given by the exact values of the percolation threshold for the Bethe-lattice with coordination number $d$.

A closer look at the perimeter polynomials (\ref{U1}-\ref{U5}) supports the hypothesis that the situation is quite similar for one-mismatch-graphs. For example, the contribution from loops to $p(4)=p^4 D_4(q)$ is given by $p^{\textrm{loop}}(4)=p^4q^{4d-4}d(d+1)/2$. Obviously, the relative weight of this term vanishes in the high dimension limit. Additional support comes from the observation that the one-mismatch base-graph $G^1_d$ contains hypercubes up to dimension $d-1$. To elucidate this property we consider the following example in $d=4$. We construct a hypercube of dimension $3$ applying the one-mismatch rule allowing for a mismatch in one of the, e.g., last three bits. Starting for instance at the origin $(0,0,0,0)$ this yields $(1,1,1,0)$, $(1,0,1,1)$, $1,1,0,1)$. Iteration gives the new vertices $(0,1,0,1)$, $(0,0,1,1)$, $(0,1,1,0)$, and finally $(1,0,0,0)$. These $8$ vertices are connected by the matching rules like a 3-dimensional hypercube. Then follow the arguments of \cite{Kahng,LD2}.

Consequently
the percolation threshold for one-mismatch-graphs can be approximated to
\begin{align}
 \label{pc1}
 p^{(1)}_c=1/d \textrm{,}
\end{align}
whereas for two-mismatch graphs holds
\begin{align}
 \label{pc2}
 p^{(2)}_c=1/(d+d(d+1)/2) \textrm{.}
\end{align}
Corrections to (\ref{pc1}) and (\ref{pc2}) are of order $O(d^{-2})$ and $O(d^{-3})$, respectively. From considerations of Sec. I we find that the number of links connecting vertices in $E_s$ with vertices in $E_{s+1}$ decreases with growing distance $s$. Thus, we conclude that the corrections to $p^{(1)}_c$ must be positive.

\begin{figure}[tbp]
 \begin{center}  
  \epsffile{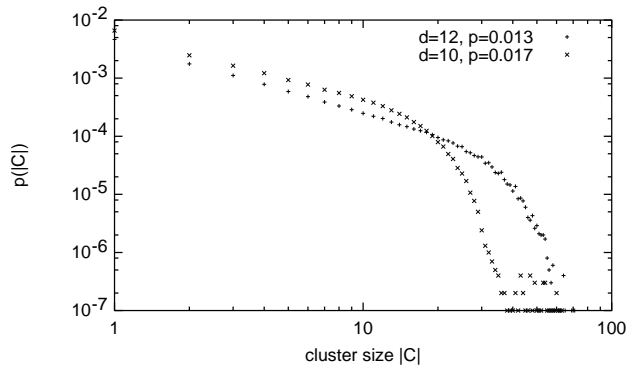}  
  \vspace*{0.5cm}
  \caption{Illustration of finite size scaling laws for the two-mismatch-graphs $G^2_{10}$ and $G^2_{12}$. Data sets are plotted on doubly logarithmic scales.}
  \label{abbu}
 \end{center}
\end{figure}

Typically for percolation problems, certain system characteristics obey scaling laws at the percolation threshold. This kind of laws reflects the statistical self similarity of clusters on different length scales in that parameter regime. Fig. \ref{abbu} displays simulation results for the cluster size distributions obtained for one-mismatch-graphs of dimensions $10$ and $14$ . Both data sets can be well described by the finite size scaling ansatz
\begin{align}
 p(|C|)=|C|^{-\tau}F(|C|/|C|^*) \textrm{ ,}
 \label{scale}
\end{align}
where the function $F(x)$ is nearly a constant for $x\ll 1$ and will be more rapidly declining than
any power law for arguments $x\gg 1$. Equation (\ref{scale}) should apply for clusters which are
neither too small nor too large. Due to the limited size of the investigated systems, very large
clusters behave in a different way than clusters of `standard' size. The transition between the laws applying to these separate cases is marked by $|C|^*$, which in turn depends on the extent of the system.

Analogous properties can be observed for base graphs defined by other matching rules. Figure \ref{abbu}
illustrates scaling behaviour for the case of graphs diced on two-mismatch-base graphs for values of $p=p^{(2)}_c$ cf. (\ref{pc2}). As a natural consequence of adding supplementary edges distances between vertices will become smaller in comparison to one-mismatch-graphs. This gives an explanation for the fact, that typical sizes $|C|^*$ are substantially smaller than those examined on one-mismatch-graphs with the same choice of $d$.

Furthermore, it can be seen in Fig. \ref{abbu} that the scaling law (\ref{scale}) even well applies to very small clusters. It is useful to define a cluster size dependent exponent $\tau_{|C|}$ by
\begin{align}
  \tau_{|C|}= -\frac{\ln \frac{p(|C|+1)}{p(|C|)}}{\ln \frac{|C|+1}{|C|}} \textrm{ .}
 \label{tau}
\end{align}
Evaluating the perimeter polynomials (\ref{U1})-(\ref{U5}), values of $\tau_{|C|}$ for small $|C|$ can be derived. Thus we obtain a sequence $\{ \tau_{|C|} \}$ which should approach the true value of $\tau$ for large $|C|$.

As a matter of fact (\ref{scale}) changes to $p(|C|)\sim |C|^{-\tau}$ in the limit of infinite
systems. Performing the limits $d\to \infty$ in Eqs. (\ref{U1})-(\ref{U5}) and (\ref{tau}), we observe that $\tau^\infty_{|C|}$ obeys
\begin{align}
 \label{tauf}
 \tau^\infty_{|C|} = \frac{1}{\ln(1+\frac{1}{|C|})}-|C|+1 \textrm{ ,}
\end{align}
which by comparison of (\ref{tau}) with (\ref{tauf}) and solving the recursion relation implies the law
\begin{align}
 p(|C|)= \frac{e^{-|C|} |C|^{|C|-2}}{|C|!}
\end{align}
for the cluster size distribution.
Equation (\ref{tauf}) leads to $\tau = 3/2$ for large clusters. The thus computed value $\tau=3/2$
is in accord with the exact result on the Bethe-lattice.

 Since we investigate a model for a biological system our main interest is devoted to large, yet finite systems. In
the sequence $\{ \tau_{|C|}\}$ our best approximation for the real value of $\tau$ is $\tau_4$. A comparison
between  values of $\tau$ computed by $\tau \approx \tau_4$ and exponents $\tau$ obtained by evaluating numerical
data for small systems suggests that the involved approximation becomes rapidly more accurate if the system size is
increased. Hence in $d=32$ we rely on $\tau_4$ and find $\tau \approx 1.5$, which gives a result that also supports
the previous assumption.

Subsequently it was our aim to find a quantity which gives an overall estimation of deviations
between random graphs on one-mismatch-graphs from those on Bethe-lattices of equal coordination number. For this purpose it appears appropriate to investigate the results of an edge elimination procedure which computes the number of bonds belonging to loops. Similar procedures have very recently
\cite{Frag} been applied to lattices.

The above aim is achieved by individually removing every edge of every cluster and calculating the number of the
connected components of the resulting graph. An edge is called fractioning if two components are obtained cutting
this edge. Clearly, only edges that belong to loops are not fractioning.

\begin{figure}[tbp]
 \begin{center} 
  \epsffile{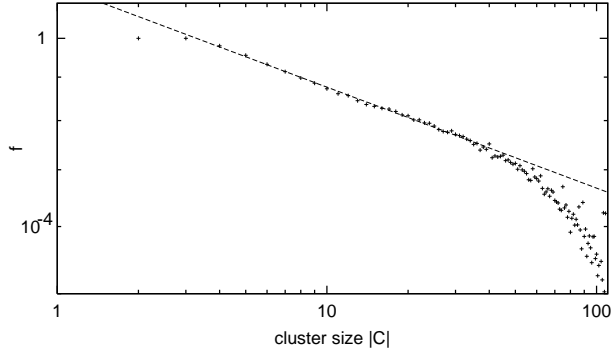}  
  \vspace*{0.5cm}
  \caption{Simulation results for the fragmentation rate $f$ on $G^1_{10}$ depending on the
cluster size $|C|$ at the numerically determined $p_c\approx 0.112$. Data are plotted on doubly logarithmic scales. Fitting a
power law (dashed line) yields $\lambda \approx 0.12$ to be compared with $\lambda \approx 0.11$ from Eq. (\ref{asym}). }
  \label{fragrate} 
 \end{center}
\end{figure}

Then the ratio of fractioning bonds $f(C)$ to the overall number of bonds $b(C)$ of a cluster $C$, $f(C)/b(C)$, is some indication for the importance of loops within the structure of $C$. Distinguishing clusters according to their size $|C|$ we computed the mean fractioning ratio
\begin{align}
f_{|C|}=<f(C')/b(C')>_{|C'|=|C|} \textrm{ .}
\label{meanfrac}
\end{align}
Except for very small clusters with trivial structure we expect a finite size scaling law
\begin{align}
 f_{|C|}=|C|^{-\lambda}\hat{F}(|C|/|\hat{C}^*|)
\end{align}
for the fragmentation rate $f$ at the percolation threshold, the validity of which is illustrated in Fig. \ref{fragrate}.

Similar to $\tau_{|C|}$ a cluster size dependent exponent $\lambda_{|C|}$ may be defined. The only 4- and 5-clusters containing non--fragmenting bonds are the 4-loop (no fractioning bond) and the 4-loop with a tail ($1/5$ of all bonds fractioning) which give the contributions $p^{\textrm{4-loop}}(4)$ and $p^{\textrm{tailed 4-loop}}(5)=5/2p^5q^{5d-7}(d-1)d(d+1)$ to $p(4)$ and $p(5)$, respectively. Making use of (\ref{U4}) and (\ref{U5}) we then find
\begin{align}
 f_{4} = \frac{p(4)-p^{\textrm{4-loop}}(4)} {p(4)} = \frac{1} {1+ \frac{1} {4(d-1)(1+1/(3q))} }
\end{align} and
\begin{align}
 f_{5} = \frac{p(4)-p^{\textrm{tailed 4-loop}}(5)} {p(4)}+\frac{1}{5} \frac{p^{\textrm{tailed 4-loop}}(5)}{p(4)} = 1-\frac{4}{5} \frac{1}{1+1/q+(d-2)(1+q+1/(12q^2))} \textrm { .}
\end{align}.
Finally, inserting $p \approx p_c^{(1)}=d^{-1}+O(d^{-2})$ a rough estimate for $\lambda$ is $\lambda_4=(\ln f_5 - \ln f_4)/\ln (5/4)$ which gives  

\begin{align}
  \lambda_4 (d)\approx \frac{1}{\ln (5/4)}\left( \frac{393}{2\cdot 10^3} d^{-1} + \frac{4029120}{8\cdot 10^6} d^{-2} + O(d^{-3})\right) \textrm{ .}
 \label{asym}
\end{align}

For $d=10$ we have $\lambda \approx 0.11$ in very good agreement with the numerically obtained value $0.12$, cf. Fig. \ref{fragrate}. For $d=32$ we find $\lambda \approx 0.03$ already very close to $\lambda=0$ which holds for
Bethe-lattices. Thus the value $\lambda > 0$ measured on finite dimensional random graphs on one-mismatch base-graphs --
which is caused by a small number of loops -- should quantify the deviation from random graphs on
Bethe-lattices. It appears that the occupation probability at the percolation threshold is still small enough to apply the Bethe approximation.

Yet, random graphs at $p_c$ are not exactly Bethe--like and contain a certain fraction of loops.
Otherwise it would be impossible to distinguish subclusters according to their connectivity within any considerable connected component. This, however, is likely to be necessary, to explain the role of the central part of idiotypic networks properly
\cite{B}.

Further investigations concerning the structure of great clusters will be made in Sec. V.

\section{Density threshold and the completeness of the idiotypic network}

Hitherto, our main interest was devoted to the question at which occupation probability a great cluster starts to exist. If $p$, however, is further increased it is also imaginable that a situation occurs, where the whole system consists of one great cluster only. This is referred to as the question about the connectivity property of random graphs.

In the theory of random graphs some general results that address related problems have been derived for so called sequences of configuration spaces
\cite{Rei2,Rei1} and earlier for a different general class of random graphs in \cite{ErdRey,Palmer}. It can be shown that the sequence of one-mismatch-base graphs $\{G^1_d\}$
fulfills all requirements of these configurations spaces \cite{Ich}, due to its rather technical
nature the proof will be omitted here.

General results of \cite{Rei2,Rei1} can be applied here showing that for infinite systems there exists a threshold
\begin{align}
 p_{\textrm{conn}}=1-\lim_{d\to \infty} |G^1_d|^{-1/\gamma_d}
 \label{eq31}
\end{align}
($\gamma_d$ being the coordination number) with the property that almost all random graphs are connected for $p>p_{\textrm{conn}}$ while the set of all connected random
graphs has measure zero for $p<p_{\textrm{conn}}$.

Moreover, for sequences of configuration spaces it holds that both the connectivity and the density
treshold, i.e. the threshold for the property that there is no non-occupied site without occupied neighbours on the base graph, coincide. Using (\ref{eq31}) we find
\begin{align}
 p^{(1)}_{\textrm{conn}}=p^{(1)}_{\textrm{dense}}=1/2
\end{align}
in case of one-mismatch-graphs and
\begin{align}
 p^{(2)}_{\textrm{conn}}=p^{(2)}_{\textrm{dense}}=0
\end{align}
for two-mismatch graphs.

Density of random graphs in our model can be directly translated into biological terms. Since antibodies and
antigens are represented by the same sets of bit chains the property that there is no free site without occupied
neighbour means that every antigen is sure to encounter a complementary antibody. Hence the idiotypic network is able
to respond to any antigen. This is meant by the completeness axiom
\cite{Ueb} for the immune system.

Nevertheless it is somewhat difficult to reconcile the demands for denseness of random graphs and occurance of small
clusters at the same time. However, it seems
unlikely that completeness should be understood in this strict way. Rather it appears to be a better solution to consider the
fact, that evolution has most likely driven the idiotypic network into such a kind of arrangement, that it is able
to respond to variations of really existing antigen only. Thus we argue that completeness of the idiotypic network
does not exactly match density of random graphs, but requires only the probability for the density of the
corresponding random graph to be somewhat below $1$. So, random graphs could still comprise small clusters and be
complete.

Another paradoxon arises if two-mismatch graphs are considered. For those the connectivity and percolation
thresholds $p^{(2)}_c$ and $p^{(2)}_{\textrm{conn}}$ fall together. How could then small clusters and a seperate
great component coexist if all random graphs are connected? Yet, for the case of finite systems it is clear that
$p^{(2)}_c(d) < p^{(2)}_{\textrm{conn}}(d)$, i.e. a great cluster has to be formed first, before it can devour his competing
small rivals.

Consequently, also for the case of two-mismatch-graphs the biologically interesting regime is well defined. For
finite systems there is a range of occupation probabilities $p^{(2)}_c(d)<p<p^{(2)}_{\textrm{conn}}(d)$, where all
requirements are met. Due to the fact that both probabilities $p^{(2)}_c(d)$ and $p^{(2)}_{\textrm{conn}}(d)$ are converging
to zero in the limit $d \to \infty$ it follows that the extent of this range $|p^{(2)}_c(d)-p^{(2)}_{\textrm{conn}}(d)|$ will
also become very small for large systems.

In the next Section our interest will be shifted towards the intrinsic structure of great clusters. Therefrom further
conclusions about biologically relevant parameter regimes can be drawn.

\section{Intrinsic structure of great clusters}
Resuming results for percolation on mismatch-graphs from Sec. I-IV two phase transitions have been found to occur.
The ranges below $p_c$ (since there is no great cluster) and above $p_{\textrm{conn}}$ (since there is only one great cluster) are of no interest concerning the biological background of the model. The one great cluster--many small clusters situation between both, however, is likely to fulfill some requirements for idiotypic networks (see Sec. II). In this Section further investigations into the network structure within this parameter regime will be made.

Insights into the structure of a great cluster can be obtained from the mass distribution of this cluster $M_c(s)=\{v\in C|d(v,c)=s \}$, i.e. the information, how many vertices have a certain distance $s$ from a starting point $c$. Due to the equality of all starting points $c$ we define
\begin{align}
 M(s)=\frac{1}{|C|} \sum_{c \in C} M_c(s)
\end{align}
and consider the mean value of $M(s)$ calculated over all clusters, whose size exceeds a minimum value $0.5\times
p\times 2^d$. For the metrics $d(\cdot \, ,\cdot)$ there are two distinguished choices, viz. metrics
defined by
(\ref{metric}) allowing paths in $G=G^1_d$, i.e. $d(\cdot \,,\cdot)=d_G$, and such restricting paths to $C$ itself, i.e.
$d(\cdot \, ,\cdot)=d_C$. Since $d_C>d_G$ we confine our investigations to $d(\cdot \, ,\cdot)=d_C$ which provides a better `resolution'. In the
following we will discuss some typical cluster compositions to obtain a survey of possible conclusions from mass
distributions to cluster structures.

\begin{figure}[tbp]
 \begin{center} 
  \epsffile{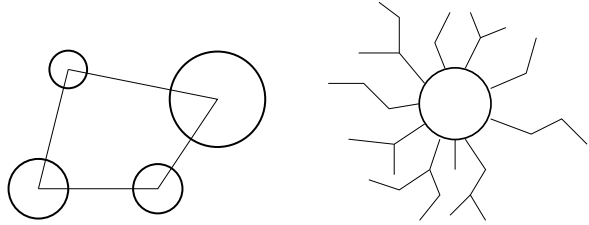}    
  \vspace*{0.5cm}
  \caption{Visualization of different cluster structures. The circles symbolize strongly connected
  subclusters (heaps). Left hand side: the cluster consists of several weakly connected heaps. Right hand
  side: The cluster contains one heap only.}
  \label{picturesque} 
 \end{center}
\end{figure}

Basically, three different situations could be imagined (for (i) and (iii) see Fig. \ref{picturesque}).

(i) A cluster could consist of some loosely clinged  high connectivity regions (heaps), whose
vertices are distinguished by a great number of connections with each other. Other vertices are
bound with relatively few links. Clearly, strong connectivity within heaps means, that vertices
belonging to them have nearly the same distance from all other vertices. Thus, the presence of many
heaps should result in a great number of local extrema, caused by `looking from all heaps' to each
other. On the other hand, loosely bound vertices will smooth the mass distribution, i.e. reduce
the sharpness of extrema.

(ii) A cluster could contain no distinguished parts at all. From (\ref{eq3}) we know that till
$s=D_{\textrm{max}}$ the
number of vertices is increasing with the distance $s$. For small occupation probabilities $p<p_{\textrm{conn}}$ (for which
not the whole base-graph has been `conquered' yet) we expect a compromise between with $s$ declining probabilities
that a vertex of that cluster has distance $s$ from an origin $v$ and the increasing number of vertices with $s$.
Consequently, the mass distribution should exhibit one maximum. For large $p>p_{\textrm{conn}}$ mass distributions can simply be derived by multiplying (\ref{eq3}) with $p$.

(iii) As a special case of (i) a cluster could be made of one heap and a certain fraction of loosely bound vertices.
Then up to two maxima, caused by the heaped vertex concentration and the competing tendencies (see (ii)),
respectively, are likely to occur. The sharpness of both maxima will depend on the fraction of the loosely bound
vertices. Thus, in case of a large proportion of those vertices both extrema could be smoothened to just one broad
maximum.

\begin{figure}[tbp]
 \begin{center} 
 \epsfxsize 8cm
 \epsfysize 5cm
 \epsffile{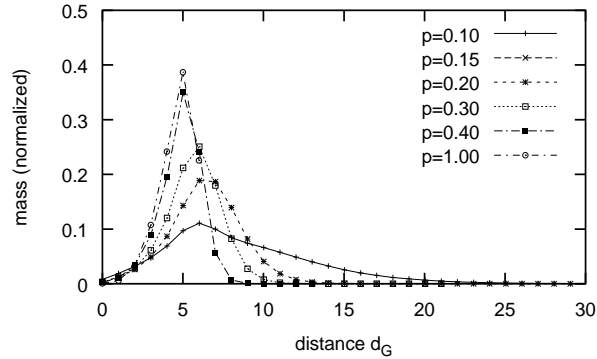} 
 \vspace*{0.5cm}
 \caption{Simulation results for the normalized mass distribution on $G^1_9$ for varying probabilities $p=0.1,0.15,0.2,0.3,0.4$ and the results of Eq. (\ref{eq3}) for $p=1.0$ (In order to make results for different $p$'s comparable on the same scale masses have been divided by the cluster size $|C|$). Points have been connected to guide the eyes.}
 \label{simpic8}
 \end{center}
\end{figure}

Fig. \ref{simpic8} shows simulation results for the normalized mass distribution for different values of $p$.
Clearly, all distributions are marked by only one maximum, whose sharpness is increasing with growing values of $p$. In the vicinity of the density threshold clusters are already stretched over the whole extent of the base graph.
Hence, for $p=0.4$ slightly below $p_{\textrm{conn}}=0.5$ the curve of the mass distribution looks similar to the exactly known distribution for the fully occupied base-graph given by Eq. (\ref{eq3}), cf. also Fig. \ref{pic1}.

More interestingly, for small $p$'s in the vicinity of the percolation threshold relatively broad maxima occur, which could be an indication for cluster structures as described in (iii).

To prove the validity of this hypothesis we have applied the edge-elimination algorithm (see Sec. II) to great
clusters. As a slight extension of the described procedure all fractioning edges are removed, thus splitting a
cluster $C$ into the sequence of its doubly connected components, or heaps in the above sense, $(C_1,...,C_t)$.
Obviously, vertices belonging to doubly connected components are distinguished by their strong cliquishness in
comparison to other vertices. Thus, the number of such components $t$ should allow to distinguish between the situations
(i)-(iii).

\begin{figure}[tbp]
 \begin{center} 
 \epsffile{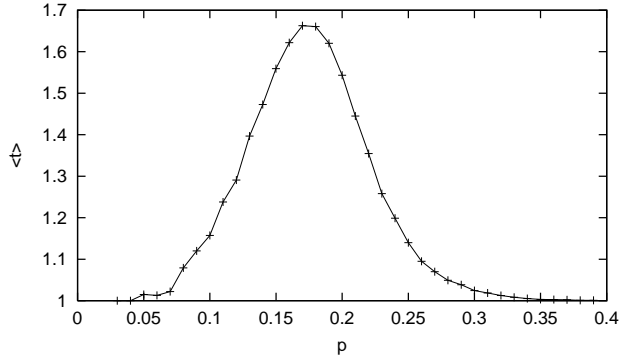} 
 \vspace*{0.5cm}
 \caption{Mean value $\langle t\rangle $ of the resulting great fragments after application of the edge-elimination procedure for $G^1_9$ depending on the occupation probability $p$.}
 \label{teilmw}
 \end{center}
\end{figure}

Fig. \ref{teilmw} shows simulation data for the mean value $\langle t\rangle$ of the resulting non-trivial parts
after edge-elimination depending on $p$. The distribution is marked by one maximum, which is again
a consequence of two competing tendencies. For small values of $p$ clusters are generally not
doubly connected, but increasing $p$ leads to a larger proportion of vertices that belong to loops.
On the other hand, there is a tendency that loops get connected by loops, i.e., seperate doubly
connected cluster components are growing together. Thus, for large $p$ almost every cluster will
consist of one doubly connected component only.

\begin{figure}[tbp]
 \begin{center} 
 \epsffile{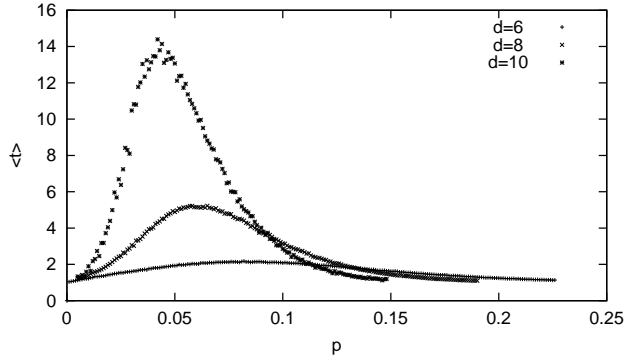}
 \vspace*{0.5cm}
 \caption{Mean value $\langle t\rangle $ of the resulting great fragments for the two-mismatch-graphs $G^2_6$, $G^2_8$ and
 $G^2_{10}$ depending on $p$. Except a shift towards smaller values of $p$ the results are similar to those on one-mismatch-graphs}
 \label{teilmw1}
 \end{center}
\end{figure}

Results of Fig. \ref{teilmw} show also that the maximum is reached slighly above the percolation
threshold $p_c$. Then, the number of doubly connected parts $t$ is rapidly declining till it
asymptotically approaches $t=1$ for $p\to 1$. We argue that this behaviour is caused by one great
doubly connected component which occurs first for some $p^>_\approx p_c$ and then gradually
incorporates all other doubly connected parts. In principle, these results do also apply to
two-mismatch graphs (see Fig. \ref{teilmw1}).

Comparing this with the above described scenario of cluster structures, we can thus state that there is a range of values of $p^>_\approx p_c$ where (iii) applies, i.e. clusters are made of one great doubly connected component (including several small ones) and a set of other loosley bound vertices. Accordingly, within this range of $p$ we define two subsets of great clusters $C$, namely the great doubly connected component $B=C^{(2)}$ and the complementary set $P=C-B$.

As already discussed in Sec. I, the central part of idiotypic networks (corresponding to great
clusters $C$) should contain strongly and weakly connected distinguished subsets. From the previous
analysis it becomes clear, that there is a choice of the only parameter $p$, where bit-string
models can exactly reproduce such a situation.

\section{conclusions}

We investigated statistical properties of a bit-string based model for idiotypic networks and compared typical
architectures of the thus defined network with axioms and hypotheses for idiotypic networks from theoretical biology. Before introducing randomness
we undertook an analysis of the underlying base graphs to show a major difference to standard
percolation problems which bases on the way how the size of the system is increased.

In the following the expressed antibody repertoire was identified with graphs, created by randomly occupying a matching
rule defined base structure (base graphs). Concepts of percolation theory have been applied in order
to determine the percolation threshold.

The immune system is a very large, yet finite system. Consequently, finite size corrections have to
be taken in account. Series expansion techniques allowed the calculation of two critical exponents (for finite systems and in
the limit $d\to \infty$) that characterize the scaling behaviour. A regime of values for the parameter $p>p_c$ has been
found where random graphs consist of many small clusters and one great connected component. For
choices $p^>_\approx p_c$ our
model reproduces the peripheral/central part conception for idiotypic networks.

The translation of the notation of the completeness of idiotypic networks into the language of graph theory 
allowed the determination of an upper threshold $p_{\textrm{dense}}$ for the occurance of the many small
clusters--one great cluster situation. Furthermore, relying on general results of the theory of random graphs (for
so-called sequences of configuration spaces) we calculated the -- in this case coinciding -- thresholds for the
density ($p_{\textrm{dense}}$) and the connectivity ($p_{\textrm{conn}}$) properties.

Furthermore, we developed techniques to obtain additional information about the structure of great clusters. Analysing
random graphs for $p^>_\approx p_c$ great clusters can be decomposed into two subsets of vertices with different
binding properties, namely groups of doubly connected vertices (backbone) and loosely linked vertices (peripheral
part of the great cluster).

Thus, bit-string models are suited to describe a hierarchy of connectivity levels, that really existing idiotypic
networks are also expected to exhibit. Our results support to some extent idealized architectures
used in a mean field type model to describe the dynamics of the central part of the immune system
\cite{B}. Contrary to other model approaches
\cite{NW92b,WBP,NW92a,ANP,Richter} for topologies of idiotypic networks our
simple few parametric bit-string model produces a non-trivial seemingly realistic network topology, without
assuming a priori distinguished vertex groups.

\begin{appendix}
\section{Renormalization group approach}
In this appendix we want to present a method to approximate the connectivity threshold for
one-mismatch-graphs. The extension of ideas of the renormalization group theory could also be applicable to more complicated cases which do not allow an exact treatment. 

Our approach bases on the idea of the renormalization of small cells \cite{Ren1} which employs the self
similarity of the system on different length scales at the percolation threshold. A condition to
apply this procedure is that an appropriate grouping of sites on the original lattice leads to a 
renormalized lattice with the same symmetry properties as the original one. Here, treating not a
real space lattice but functional networks, we adapt the idea of the renormalization group theory in
the following way. We find a transformation ${\cal R}$ which, by grouping of vertices, leads from the
one-mismatch base graph $G^1_{d+2}$ in dimension $d+2$ to a graph ${\cal R}(G^1_{d+2})$ which is equivalent
to the base graph $G^1_d$ in dimension $d$, ${\cal R}(G^1_{d+2})\simeq G^1_d$, thus allowing a systematic
reduction of the degrees of freedom.

If a -- for
finite systems appropriately defined -- threshold property, namely $p_{\textrm{conn}}(d)$, converges to a
certain value for $d\to \infty$, differences between $p_{\textrm{d}}$ and $p_{\textrm{d}-2}$ must be
small and will disappear in the limit of infinite systems. Consequently, if a grouping of vertices
to super vertices on a graph $G$ yields a renormalized graph ${\cal R}(G)$ of the same type, percolation
thresholds on both graphs will be the same. Thus, we replace the term symmetry (as it is applied to
lattices) by equivalency of graphs.

\begin{figure}[tbp]
 \begin{center}
 \epsffile{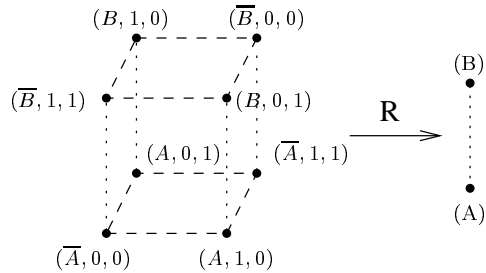}
 \vspace*{0.5cm}
 \caption{Illustration of the renormalization procedure applied to a pair of connected 4-loops on
 $G^1_{d+2}$ leading to a pair of linked super vertices on $G^1_d$. If $d_H(B,\overline{A})=0$ the
 dotted edges correspond to inversion links, if $d_H(B,\overline{A})=1$ to one-mismatch links.}
 \label{anh1}
 \end{center}
\end{figure}

We encode an arbitrary vertex of the base graph $G^1_{d+2}$ by $(A, b_0, b_1)$ where $A$ is a bit
chain of length $d$ and $b_0$, $b_1$ are single bits. Every vertex on $G^1_{d+2}$ belongs to a
unique 4-loop $\{(A, b_0, b_1), (\overline{A}, \overline{b_0}, b_1), (A, \overline{b_0},
\overline{b_1}), (\overline{A}, b_0, \overline{b_1})\}$ connected by one-mismatch links. The
renormalization ${\cal R}$ replaces this 4-loop by the `super'vertex $(A)$ on ${\cal R}(G^1_{d+2})$ (of course the
choice of $(\overline{A})$ leads to identical results). If two vertices $A$ and $B$ are connected by
an inversion (one-mismatch) link on $G^1_d$ the vertices of the corresponding 4-loops on $G^1_{d+2}$
are connected by 4 inversion (one-mismatch) links, too, cf. Fig. \ref{anh1}.

We thus define a renormalized graph ${\cal R}(G^1_{d+2})$ composed of the above explained vertices and edges. It
follows directly from our construction that ${\cal R}(G^1_{d+2})$ is equivalent to $G^1_d$.

To describe random graphs
we apply the following majority rule: a super vertex on $G^1_d$ is considered as occupied if at
least two connected vertices of the corresponding 4-loop on $G^1_{d+2}$ are occupied,  i.e. this
4-loop is said to percolate. On the basis of this
rule we obtain
\begin{align}
\label{ren1}
p'=4p^2(1-p)^2+4p^3(1-p)+p^4 \textrm{ ,}
\end{align}where $p'$ denotes the (renormalized) probability that vertices of $(G)$ are occupied if
vertices on $G$ were diced with probability $p$.

Calculating fixed points of eq. (\ref{ren1}) we obtain $p^*=0$, $(3^+_-\sqrt{5})/2$, and $1$.
Thus, as the only unstable solution in $[0,1]$ we have $p^*=(3-\sqrt{5})/2$ as a first
approximation for the connectivity threshold.

At this place one could pose the question whether only 4-loops are suited as a renormalization cell.
Indeed, it is also possible to summarize successive renormalization steps to a single
great one, ordinary hypercubes of even dimension $k<d$ form suitable cells as well. On the other
hand, condensing odd dimensional hypercubes to super vertices does not lead to renormalized graphs
${\cal R}(G)$
that are equivalent to other base graphs of the sequence $\{ G^1_d\}$. Altogether, this seems to be
an effect of differences between even and odd-dimensional base graphs (see Sec. II).

Of course, renormalizing small cells entails some approximation. Grouping together vertices and
applying a majority rule to occupy the renormalized vertices, situations can arise where clusters on
the original lattice are cut or new clusters are formed \cite{Ren1}.

There are different approaches to improve the involved approximations. One possibility suggested in
\cite{Ren2} which leads to exact results in the limit of very large cells is to summarize some
elementary cells to one large cell of size $z$. This large cell will then be occupied if all elementary cells
are occupied and connected, i.e. are said to percolate. Since 'renormalization faults' are
essentially due to cell surface effects improvements produced by the above method emanate from the declining
surface-volume ratio of large cells. For one-mismatch graphs it holds, however, that the surface
size of a cell $s(z)$ depends logarithmically on $z$, viz. $s(z)\approx d+1-\log_2 z$ leading to
only slight improvements with increasing $z$.

Using larger elementary cells consisting of two coupled 4-loops we obtain
\begin{align}
 \label{ren2}
 (p')^2=p^8+8p^7q+24p^6q^2+32p^5q^3+12p^4q^4 \textrm{ ,}
\end{align}
using four coupled 4-loops yields
\begin{align}
 \nonumber
(p')^4&=p^{16}+16p^{15}q+112p^{14}q^2+448p^{13}q^3+1120p^{12}q^4+ \\ \label{ren3}
 &+1792p^{11}q^5+1776p^{10}q^6+1008p^9q^7+180p^8q^8
\end{align}
where $q=1-p$. As fixed points $p'=p$ of (\ref{ren2}) and (\ref{ren3}) we determined numerically
$p\approx 0.41$ and $p\approx 0.39$, respectively. The difference of both values to $p_{conn}=0.5$
may be a consequence of the above mentioned slow convergence.

More interestingly, our renormalization procedure of replacing hypercubes of dimension $k$ by
super vertices is without modifications applicable to ordinary hypercubes as well. This gives an
additional argument that the connectivity thresholds of both sequences of graphs are equal which can
also be verified by evaluating eq. (\ref{eq31}).

\end{appendix}

\pagebreak[8]

\begin{figure}[tbp]
 \begin{center} 
  \epsffile{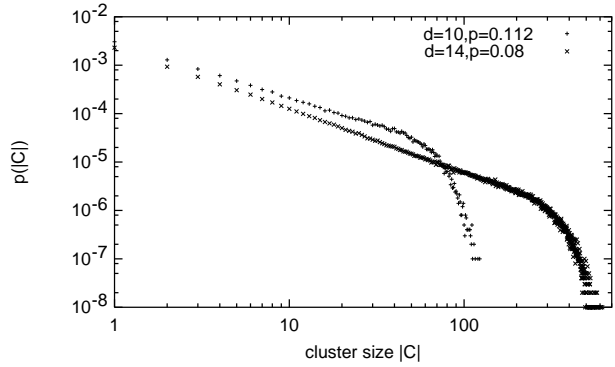}
  \vspace*{0.5cm}
  \caption{Cluster size distribution on $G^1_{10}$ and $G^1_{14}$ at
the percolation thresholds. The simulation data are plotted on doubly logarithmic scales.}
 \end{center}
\end{figure}
 
\end{document}